\documentclass[preprint,aps,prl]{revtex4-1}
\usepackage{graphicx}
\begin{document}
\title{
Lock-in transition of charge density waves in quasi-one-dimensional conductors:\\ reinterpretation of McMillan's theory}
\author{Katsuhiko Inagaki}
\affiliation{Department of Physics, Asahikawa Medical University, Midorigaoka Higashi 2-1, Asahikawa, 078-8510, Japan}
\author{Satoshi Tanda}
\affiliation{Division of Applied Physics, Graduate School of Engineering, Hokkaido University, Kita 13 Nishi 8, Kita-ku, Sapporo 060-8628, Japan}

\begin{abstract}
We investigated the lock-in transition of charge density waves (CDWs) in quasi-one-dimensional conductors, based on McMillan's free energy. The higher-order umklapp terms play an essential role in this study. McMillan's theory was extended by Nakanishi and Shiba in order to treat multiple CDW vectors. Although their theories were aimed at understanding CDWs in quasi-two-dimensional conductors,
we applied them to the quasi-one-dimensional conductors, including K$_{0.3}$MoO$_3$, NbSe$_3$, and $m$-TaS$_3$, and confirmed its validity for these cases. Then we discussed our previous experimental result of $o$-TaS$_3$, which revealed the coexistence of commensurate and incommensurate states.  We found that the coexistence of multiple CDW vectors is essential for the lock-in transition to occur in $o$-TaS$_3$. The even- and odd-order terms in the free energy play roles for amplitude development and phase modulation, respectively. Moreover, consideration of the condition of being commensurate CDWs allowed us to relate it with that of the weak localization in random media.
\end{abstract}
\maketitle

\section{I. Introduction}
Lock-in transition between incommensurate and commensurate charge density waves (CDWs) has been studied since the mid-1970s \cite{Moncton1975,McMillan1976,Bak1976,Nakanishi1977,Suits1980,Roucau1980,Chen1981,Roucau1983,Tanda1984,Fleming1984,Tanda1985,Fleming1985,Inagaki2008,Sakabe2017}.
It is induced by the coupling of a lattice periodicity with a charge density wave.
The transition is often accompanied with the formation of a discommensuration lattice between commensurate and incommensurate phases.
Occurrence of discommensuration was predicted by theory \cite{McMillan1976} and found in quasi-two-dimensional conductors, e.g. in 2H-TaSe$_2$ \cite{Suits1980,Chen1981} and 1T-TaS$_2$ \cite{Tanda1984,Tanda1985,Sakabe2017}, both of which are typical quasi-two-dimensional conductors with Peierls transition.
In contrast, the lock-in transition of CDWs in quasi-one-dimensional conductors remains unsubstantiated, although several experiments were reported \cite{Roucau1980,Roucau1983,Fleming1984,Fleming1985}.
We previously performed a synchrotron x-ray study in $o$-TaS$_3$ and suggested that the discommensuration lattice is formed when commensurate and incommensurate phases coexist \cite{Inagaki2008}.
However, this preliminary report lacked theoretical interpretation.
In this paper we review the theoretical treatments of the lock-in transition and apply them to the quasi-one-dimensional conductors, including K$_{0.3}$MoO$_3$, NbSe$_3$, and $m$-TaS$_3$. The validity of the theory is confirmed for these cases. We then go on to discuss the synchrotron data of $o$-TaS$_3$. The coexistence of the commensurate and incommensurate phases is found to be essential for the lock-in transition. \textit{The even- and odd-order terms in free energy play roles for amplitude development and phase modulation, respectively.} Moreover, consideration of the condition of being commensurate CDWs allows us to relate it with that of the weak localization in random media \cite{Bergmann1983}. This explains why quantum interference phenomena have been observed in CDWs \cite{Tsubota2012,Inagaki2016}.

\section{II. Previous experiments}
\begin{figure}[h]
\begin{center}
\includegraphics[width=0.8\textwidth]{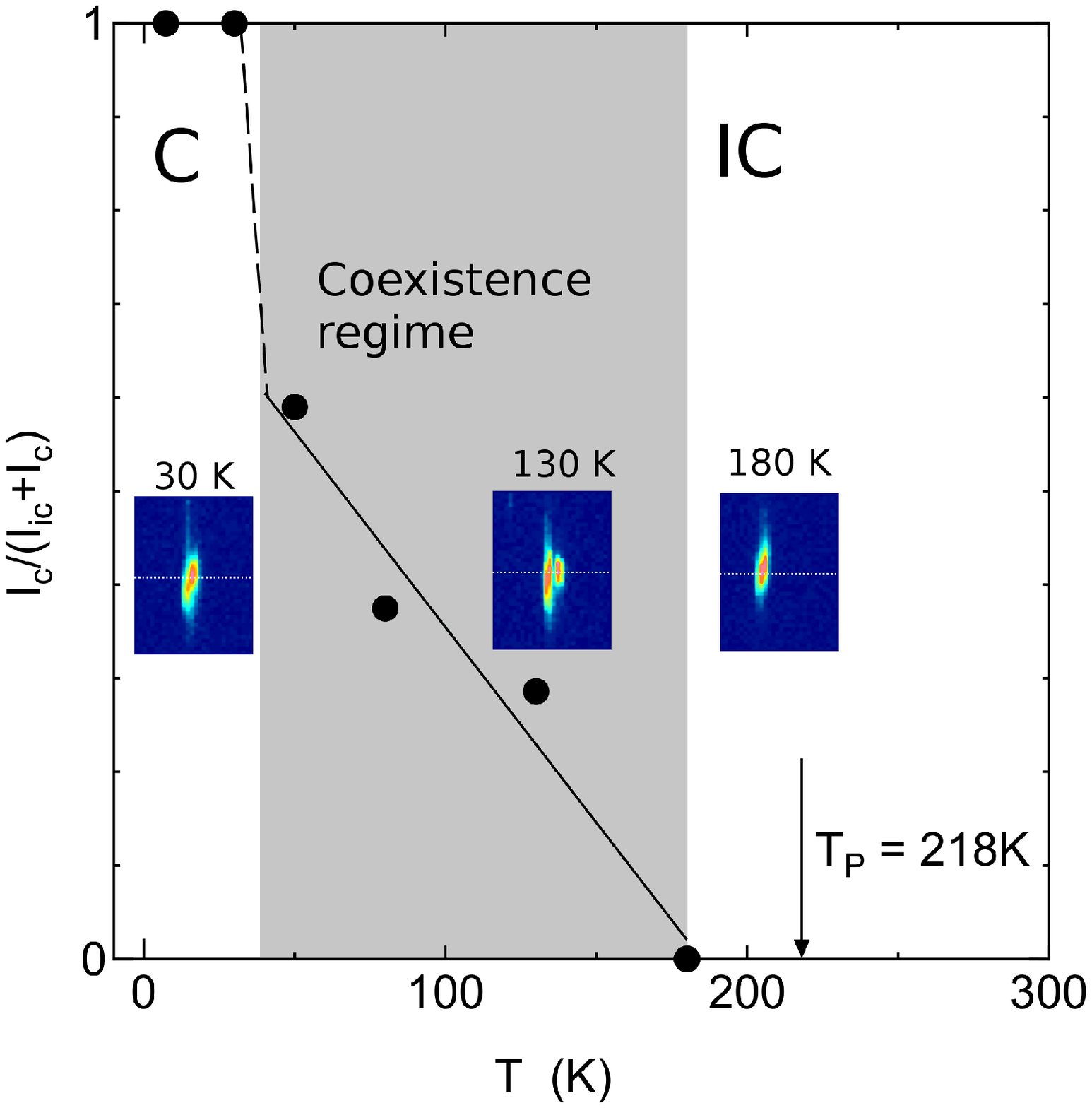}
\end{center}
\caption{
Phase diagram of $o$-TaS$_3$ deduced from
this study.
Solid circles represent
the intensity of a commensurate CDW, $I_\mathrm{c}$, normalized by the total
intensity of both the commensurate and incommensurate satellites, $I_\mathrm{ic}+I_\mathrm{c}$.
The commensurate CDW begins to develop at around 180 K (solid line).
The two $q$'s of the CDWs coexist until the system undergoes the lock-in transition at a temperature between 30 K and 50 K (broken line).
The insets are diffraction profiles of satellite peaks at 30, 130, and 180 K.}
\label{fig:phase_diagram}
\end{figure}

Let us take an overview of the lock-in transition in $o$-TaS$_3$.
This quasi-one-dimensional conductor undergoes a Peierls transition at 220 K. At the transition, all the electrons on the Fermi surfaces contribute to form the Peierls gap and the system becomes an insulator, contrary to similar materials, such as NbSe$_3$ and $m$-TaS$_3$, in which remained electrons contribute to metallic conduction even after Peierls transitions occur. Hence, by absence of remained normal electrons, $o$-TaS$_3$ is one of the most appropriate materials for CDW studies.
The first x-ray study was made by Tsusumi \textit{et al.} who determined the CDW vector of $o$-TaS$_3$ \cite{Tsutsumi1978}. This work was followed by
Roucau, who found that the CDW vector shifts
from being $0.255c^*$ (incommensurate) to $0.250c^*$ (commensurate) at low temperatures \cite{Roucau1983}. The details of the lock-in transition were revealed by use of synchrotron diffraction \cite{Inagaki2008}. By lowering the temperature, the CDW vector shifts from being incommensurate closer to commensurate;
however, it stops at $0.252c^*$. The commensurate CDW independently appears at 130 K.
In addition, coexistence of the commensurate and incommensurate CDWs was found in the temperature range down to 50 K, then the complete lock-in was observed at the lowest temperature, as shown in Fig. \ref{fig:phase_diagram} \cite{comment:phase_diagram}.
The observed diffraction pattern of coexistence of two CDWs is clearly distinguished from those in the current-induced discommensuration lattice, which induces symmetric subpeaks at both sides of the main satellite \cite{LeBolloch2008,Rojo-Bravo2016}.

The observed CDW characteristics in $o$-TaS$_3$ differ from those in other quasi-one-dimensional conductors. Blue bronze K$_{0.30}$MoO$_3$ undergoes a Peierls transition at 180 K with $k_1=0.263b^*$ \cite{Fleming1985}. By lowering the temperature to 100 K, its CDW wave vector shifts to be nearly commensurate with a slight residual incommensurability ($\sim\!0.250 b^*$). Lock-in transition to the commensurate state does not occur in this conductor (incomplete lock-in).
In NbSe$_3$ \cite{Fleming1984}, as well as $m$-TaS$_3$ \cite{Roucau1980}, there are three conducting chains, two of which contribute to form CDWs. Neither NbSe$_3$ nor $m$-TaS$_3$ exhibits lock-in transition.
The first CDW wave vector in $m$-TaS$_3$ is independent of temperature with $q_1=0.245b^*$ , whereas $q_1$ of that in NbSe$_3$ shifts from $0.245 b^*$ at 150 K to $0.241 b^*$ at low temperatures \cite{Moudden1990}.
In these conductors, the second CDW ($q_2=0.247b^*$ for $m$-TaS$_3$, $q_2=0.260b^*$ for NbSe$_3$) appears at lower temperatures.
As summarized in Table \ref{tb:transition}, it seems difficult to treat such various behaviors of CDWs in quasi-one-dimensional conductors with a simple theory, in particular, for such vector shift and complete/incomplete lock-in phenomena.

\begin{table}
\caption{Characteristics of quasi-one-dimensional CDW systems. The wave vectors of each material and its ground state are shown (commensurate, incommensurate, or nearly commensurate).}\label{tb:transition}
\begin{tabular}{c|ccc}
\hline\hline
Material & CDW 1 & CDW 2 & Ground State \\
\hline
$o$-TaS$_3$ & $0.255c^* \to 0.252c^*$ & $0.250c^*$ & C\\
K$_{0.3}$MoO$_3$ & $0.263b^* \to \ \sim\! 0.250b^*$& $-$ & NC\\
$m$-TaS$_3$ & $0.253b^*$ & $0.247b^*$& IC\\
NbSe$_3$ &$0.245b^* \to 0.241b^*$ & $0.260 b^*$ & IC\\
\hline\hline
\end{tabular}
\end{table}

\section{III. Model and Results}
Theoretically, the lock-in transition in quasi-two-dimensional conductors has been discussed, initially by McMillan \cite{McMillan1976}, and followed by Nakanishi and Shiba \cite{Nakanishi1977}. Their treatment is based on free energy with higher-order umklapp terms.
McMillan's free energy has the following form:
\begin{equation}
F_1=F^0\int d^2 s[-|\phi|^2-\beta Y Re(\phi^3)+\frac{1}{2} |\phi|^4+\beta|\vec\nabla\phi +i\phi|^2+\gamma|-\vec{q_1} \times\vec\nabla\phi |^2 ]
\label{eq:McMillan}
\end{equation}
where $\phi$ is a phase of CDW defined as $\psi=\psi_0 e^{i{K}\cdot{r}/3}\phi({r})$, and $\psi$ is a complex order parameter.
Here the coefficients $F^0$, $\beta$, $Y$, and $\gamma$ were the same as those defined in the original literature\cite{comment:coefficients}.
Equation (\ref{eq:McMillan}) was derived for commensurability index, namely, the ratio of CDW and lattice periodicities, $M=3$.
The commensurability energy originates from the third-order umklapp term, proportional to the coefficient $\beta$. Though McMillan's discussion aimed to understand the behavior of quasi-two-dimensional conductors, e.g., 2H-TaSe$_2$, it also includes quasi-one-dimensional cases.
To apply their theories to our case, $M=4$, we should know what happens in McMillan's free energy.
By substituting $\psi=\psi_0 e^{i{K}\cdot{r}/4}\phi({r})$ for the order parameter,
a simple calculation gives a result similar to Eq. (\ref{eq:McMillan}); however, it lacks the $\beta Y Re(\phi^3)$ term, because the umklapp term becomes fourth-order in this case, namely, proportional to $|\phi|^4$.
In contrary to the $M=3$ case, this calculation provides an unfamiliar result. \textit{The umklapp term gives no energy gain} if a phase modulation $\phi=e^{-i\theta(x)}$ alone is considered as in McMillan.
His calculation for the case $M=3$ provided the free energy as $F^0(-\frac{1}{2}+\beta(1-Y))$. A first-order lock-in transition takes place at the point $Y=1$. On the other hand, from our calculation for the case $M=4$, the free energy of the commensurate state is a constant value $F^0(-\frac{1}{2}+\beta)$, which is always larger than that of incommensurate state $F^0(-1/2)$.
This explains the absence of the lock-in transition in the charge density wave of blue bronze, whose CDW vector becomes nearly-commensurate at low temperatures.
On the contrary, the origin of the CDW vector shift from $0.263b^*$ to $\sim 0.250b^*$ remains unsolved. We will discuss this issue later.

Nakanishi and Shiba's extension of McMillan's theory covers the systems with multiple CDW vectors \cite{Nakanishi1977}. They treated the lock-in transition of a two-dimensional conductor 1T-TaS$_2$, whose nesting vectors $k_i\ (i=1,2,3)$ satisfy a relation $3k_i-k_j=G_i$, where $G_i$ are reciprocal vectors, leading to the commensurability energy through the fourth term of umklapp processes. Also, after a simple calculation, this fourth-order term is found to give the energy gain only when coefficients of the nesting vectors in such a relation are odd numbers (1 or 3) for combining them to the reciprocal vector. This explains the absence of the lock-in transition in NbSe$_3$ and $m$-TaS$_3$, both of which have the nesting vectors satisfying $2k_1+2k_2=G$.

\begin{figure}[h]
\begin{center}
\includegraphics[width=0.8\textwidth]{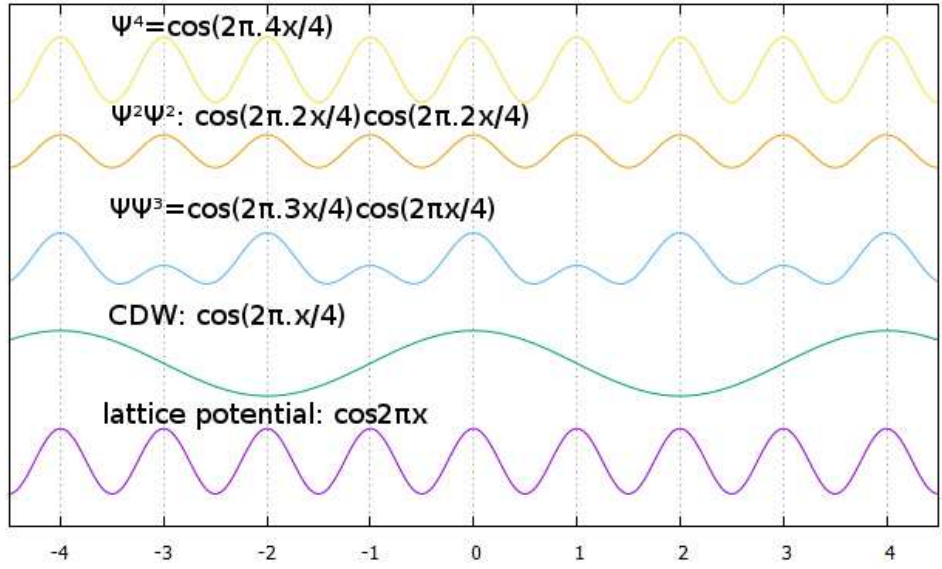}
\end{center}
\caption{Visualization of the fourth-order umklapp terms: $\psi^4$, $\psi^2\psi^2$, and $\psi\psi^3$; charge density wave of commensurability index $M=4$, and lattice potential (from top to bottom).}
\label{fig:umklapp}
\end{figure}

As shown above, even-order processes in the free energy develop the amplitude, while odd-order processes induce phase-related phenomena. Figure \ref{fig:umklapp} shows whether the fourth-order umklapp terms couple to the lattice periodicity or not. The (2,2) case, namely, $2k_1+2k_2=G$, which is satisfied in NbSe$_3$ and $m$-TaS$_3$, provides the same potential modulation as that in blue bronze. Therefore, the absence of lock-in transition in these conductors is found to be of the same origin. On the other hand, the (1,3) case provides sufficient contribution to the lock-in transition also in quasi-one-dimensional conductors. This case was discussed to explain the lock-in transition of an organic conductor, TTF-TCNQ \cite{Bak1976}.

Now we will apply these theoretical considerations to our experimental results.
CDWs in $o$-TaS$_3$ were not assumed as those in multiple chains, such as NbSe$_3$ and $m$-TaS$_3$. However, the coexistence of commensurate and incommensurate CDWs in $o$-TaS$_3$, as shown in Fig. \ref{fig:phase_diagram}, suggests this possibility.
By lowering the temperature, CDWs split into two kinds: commensurate and incommensurate ones. The commensurate CDW vector appears at $k_c=0.250c^*$ from even-order terms in the free energy, whereas the incommensurate CDW vector remains at $k_{ic}=0.252c^*$.
The fourth-order umklapp term, which satisfies $k_c+3k_{ic}\simeq G$, couples to the lattice periodicity and obtains commensurability energy. At a temperature between 50 K and 30 K, a transition may occur, allowing the system to be complete lock-in. This scenario perfectly explains our synchrotron data \cite{Inagaki2008}.
The incommensurate phase in the coexistence regime may have discommensurations, as discussed in the previous report, because a discommensuration state is energetically preferable to incommensurate CDW, according to McMillan \cite{McMillan1976}. In addition, the transition temperature coincides with that of occurrence of glasslike behavior \cite{Staresinic2002}. This behavior can be understood as a result of the lock-in transition, which freezes global motion of the CDWs.

\section{IV. Discussions}
Our discussion does not rule out the possibility for the generation of an individual discommensuration, namely soliton, in commensurate CDWs. According to Bak and Emery \cite{Bak1976}, a sinusoidal potential in CDW leads to the sine-Gordon equation, whose solution includes a phase soliton with the charge $e/M$. Moreover, such a sinusoidal modulation of potential can be derived only by commensurability \cite{Lee1974}.
This agrees with previous experimental results, including the discrepancy between longitudinal and transverse conductivity at low temperatures \cite{Takoshima1980}, the existence of unexpected carriers \cite{ZZ2006}, and the nonlocal transportation \cite{Inagaki2010}.

According to the microscopic theory \cite{Lee1974}, the sinusoidal potential in commensurate CDWs is rooted in the condition
\begin{equation}
\epsilon_{k+MQ}=\epsilon_k,
\label{eq:Lee}
\end{equation}
where $\epsilon_k$ is the energy of momentum $k$, and $Q=2k_F$ is a CDW vector. Equation (\ref{eq:Lee}) means that the sum of each vector equals the reciprocal vector, i.e., $MQ=G$, and the energy of an electron-hole pair conserves after it is interacted $M$ times by the CDW momentum of $2k_F$. This leads to the phase dependence of the gap energy as $2|\Delta|^M(\cos M\phi-1)$ \cite{LRA}.

\begin{figure}[h]
\begin{center}
\includegraphics[width=0.8\textwidth]{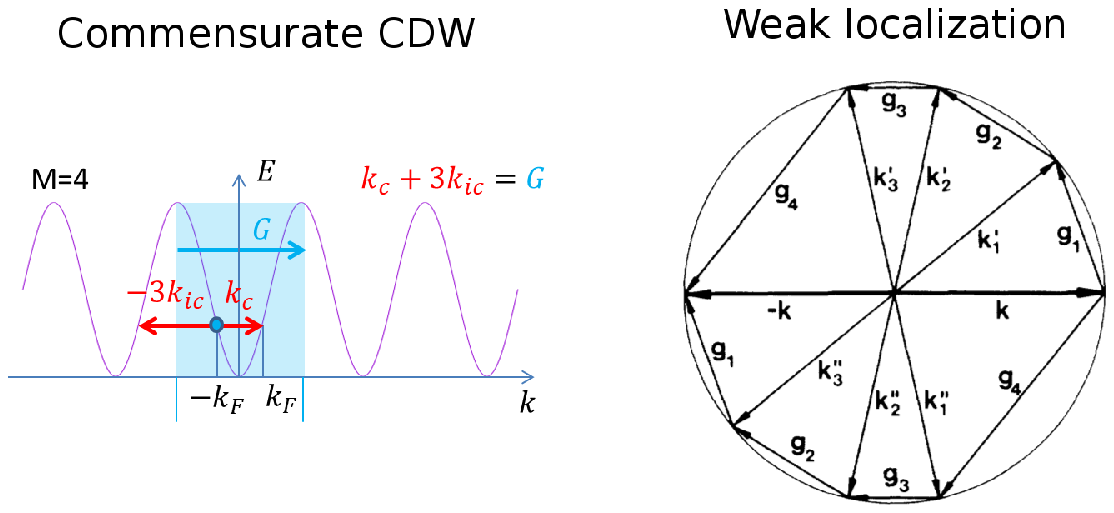}
\end{center}
\caption{Umklapp process in $M=4$ CDW, responsible to commensurability (left).  Schematic diagram of Bergmann's condition  (right) \cite{Bergmann1983}.}
\label{fig:wlr}
\end{figure}

If a system is purely one-dimensional, Eq. (\ref{eq:Lee}) merely provides $M$'th order of umklapp process, whereas in two-dimensional systems, another interpretation becomes possible as follows: it is similar to that of Anderson localization, in particular, in the weak localization regime \cite{Bergmann1983}. Anderson localization results from self-interference of a wave function by multiple elastic scattering in random media. Bergmann's condition for the localization to occur has a form $\sum_i g_i=0$, where $g_i$ denotes scattering vectors by impurities. It should be noted that a moment in the lattice can stay in any arbitrary Brillouin zone.
Since all the scattering processes are elastic, the energy of the wave function conserves. Therefore, by considering Bergmann's condition in substitution of $Q$ for $g_i$, one may obtain Eq. (\ref{eq:Lee}), as shown in Fig. \ref{fig:wlr}.

This interpretation agrees with previous experimental results in $o$-TaS$_3$.
At low temperatures, the system undergoes complete lock-in, as shown in Fig. \ref{fig:phase_diagram}, where quantum interference phenomena were discovered in $o$-TaS$_3$ \cite{Tsubota2012,Inagaki2016}. In particular, the localization phenomenon in the commensurate state suggests that CDWs have a two-dimensional correlation over the $b$-$c$ plane, and the closed path of CDW trajectory plays a crucial role \cite{Inagaki2016}.

Finaly, here we will mention a limitation to our discussion. The lock-in energy has been found to relate with the odd-order terms in McMillan's free energy. By applying this to quasi-one-dimensional conductors with $M=4$, most of characteristics summarized in Tabel \ref{tb:transition} are explained within this framework, except for the vector shift observed in blue bronze.
One plausible explanation is the excitation of soliton and antisoliton pairs \cite{Artemenko1981}.
Each excitation of the soliton and antisoliton pair has been observed as a discrete step \cite{Zybtsev2010}.
Since similar steps have also been observed in $o$-TaS$_3$ \cite{Zybtsev2016}, further investigation must be necessary to clarify the lock-in transition of CDWs.

\section{V. Conclusion}
In summary, we provide a unified view for the lock-in transition both in quasi-one- and two-dimensional conductors, based on the difference of roles between even- and odd-order terms in the free energy. The study of commensurate CDWs should be more focused, since it must contain far richer physics than previously thought.

\section*{Acknowledgments}
The authors thank K. Ichimura, K. Yamaya, T. Honma, M. Hayashi, and K. Nakatsugawa for fruitful discussions, and M. Tsubota, T. Matsuura, S. Uji, N. Ikeda, and Y. Nogami for experimental support.

\clearpage

\clearpage


\begin{thebibliography}{99}
\bibitem{LRA}
In the original paper \cite{Lee1974}, the authors insist that this result can be generalized for $M \geq 2$. However, it should be noted that the phase-dependent term becomes a different form in case $M=3$.
\end{thebibliography}
\end{document}